\documentclass[aps,prl,reprint,superscriptaddress,showpacs]{revtex4-1}

\usepackage{comment}
\usepackage{graphicx}
\usepackage{latexsym}
\usepackage{xcolor}
\usepackage{amsmath}
\usepackage{multirow}
\usepackage{placeins} 
\usepackage{float}
\usepackage{hyperref}

\graphicspath{{images/}}

\renewcommand{\vec}[1]{\mathbf{#1}}

\newcommand{\dd}{\mathrm{d}}

\newcommand{\be}{\begin{equation}}
\newcommand{\ee}{\end{equation}}
\newcommand{\bea}{\begin{eqnarray}}
\newcommand{\eea}{\end{eqnarray}}

\begin{document}

\title{Bend-Induced Twist Waves and the Structure of Nucleosomal DNA}

\author{Enrico Skoruppa}
\affiliation{KU Leuven, Institute for Theoretical Physics, Celestijnenlaan
200D, 3001 Leuven, Belgium}

\author{Stefanos K. Nomidis}
\affiliation{KU Leuven, Institute for Theoretical Physics, Celestijnenlaan
200D, 3001 Leuven, Belgium}

\affiliation{Flemish Institute for Technological Research (VITO),
Boeretang 200, B-2400 Mol, Belgium}

\author{John F. Marko}
\affiliation{Department of Physics and Astronomy, and Department of
Molecular Biosciences, Northwestern University, Evanston, Illinois
60208, USA}

\author{Enrico Carlon}
\affiliation{KU Leuven, Institute for Theoretical Physics, Celestijnenlaan
200D, 3001 Leuven, Belgium}

\date{\today}

\begin{abstract}
Recent work indicates that twist-bend coupling plays an important role
in DNA micromechanics. Here we investigate its effect on bent DNA. We
provide an analytical solution of the minimum-energy shape of circular
DNA, showing that twist-bend coupling induces sinusoidal twist waves. This
solution is in excellent agreement with both coarse-grained simulations
of minicircles and nucleosomal DNA data, which is bent and wrapped around
histone proteins in a superhelical conformation.  Our analysis shows that
the observed twist oscillation in nucleosomal DNA, so far attributed to
the interaction with the histone proteins, is an intrinsic feature of
free bent DNA, and should be observable in other protein-DNA complexes.
\end{abstract}

\maketitle

{\sl Introduction --} Elastic models of DNA have been a key tool
for understanding the response of the double helix to applied
stresses~\cite{mark15}. Such stresses are ubiquitous in cells, where
DNA is continuously being bent and twisted. For instance, in eukaryotes
about $75\%$ of the DNA is wrapped around cylindrically-shaped octamers
of histone proteins~\cite{esla16}. The 147 base pairs (bp) of wrapped
DNA sequence and the histone form the nucleosome, which represents the
lowest level of chromosomal organization.

At length scales of a few nanometers the behavior of DNA can be modeled by
a homogeneous elastic rod, with stiffness constants associated with the
different types of mechanical deformations~\cite{bust94, mark95, stri96,
moro98}. The simplest such model is the twistable wormlike chain (TWLC),
which treats bending and twist as independent deformations. However,
symmetry analysis of the right-handed, oppositely-directed-backbone
double helix indicates that there must be a coupling of bending to
twisting~\cite{mark94}. This can be understood as a consequence of the
asymmetry between the major and minor grooves of the double helix. Only a
few prior works have considered twist-bend coupling~\cite{olso98,lank03,
cole03, olso04, moha05, drvs14, nomi17, skor17}, and its effect on
equilibrium and dynamics of DNA remain largely unexplored.

Here we investigate the effect of twist-bend coupling on free DNA
minicircles and compare their shapes with X-ray crystallographic
structures of nucleosomal DNA (DNA wrapped around histones). We present
an analytical solution of the minimal energy configuration of free
minicircles which shows that twist-bend coupling induces sinusoidal twist
waves coupled to bending waves. The results are in excellent agreement
with molecular dynamics simulations of two different coarse-grained DNA
models~\cite{ould10}: one with symmetric grooves and one with grooves of
unequal widths. Only in the latter twist waves are observed, in agreement
with the symmetry argument of Ref.~\cite{mark94}. The nucleosomal DNA
shape obtained from averaging $145$ available crystal structures displays
twist waves quantitatively matching the predictions of our simple theory
for free DNA. While several studies in the past analyzed oscillations
in twist in nucleosomal DNA, this was usually attributed to interations
with the underlying histone proteins~\cite{esla16}. Our work shows that
twist waves are a general feature of bent DNA and that similar results
should be observable for other protein-DNA complexes.

\begin{figure}[b!]
\centering\includegraphics[width=\linewidth]{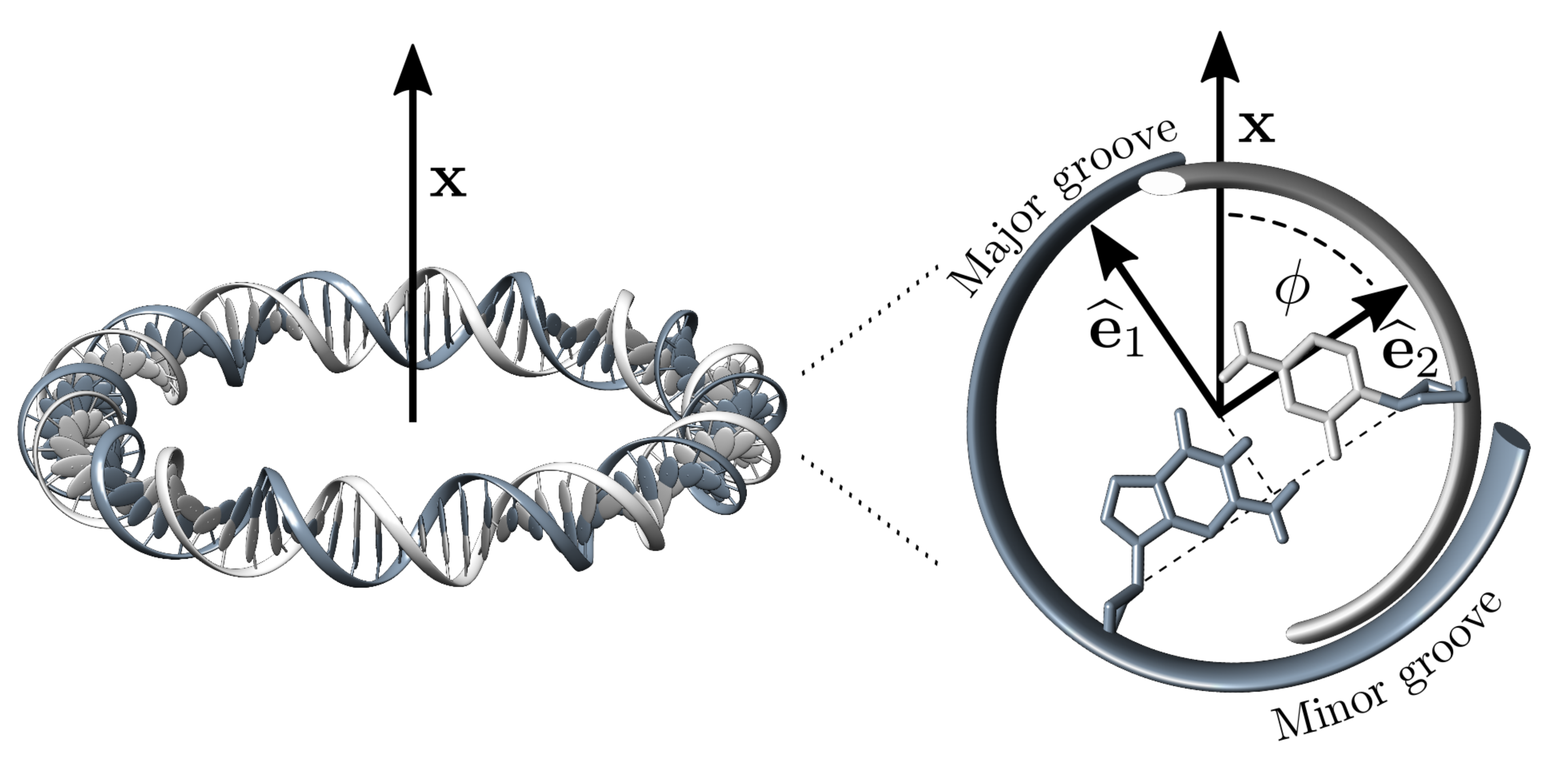}
\caption{Left: Schematic view of a DNA minicircle lying on a plane
orthogonal to a vector $\vec{x}$. Right: Zoom-in of a
cross-section of the double helix showing the unit vectors $\vec{\widehat
e}_1$ and $\vec{\widehat e}_2$ (the tangent vector $\vec{\widehat e}_3 =
\vec{\widehat e}_1  \times \vec{\widehat e}_2$ points inside of the
page). In an ideal fully-planar circle $\vec{x}$ lies on the
plane spanned by $\vec{\widehat e}_1$ and $\vec{\widehat e}_2$. $\phi$
is the angle between $\vec{\widehat e}_2$ and $\vec{x}$.}
\label{fig:scheme}
\end{figure}

{\sl Theory and Energy Minimization --}  
Following prior work~\cite{mark94}, we describe the double helix
centerline using a space curve in arc-length parameterization,
with coordinate $s$ running from $0$ to the total DNA length $L$;
we thus treat the double helix as inextensible, which turns out
to be appropriate for our purposes.  Along the curve we define an
orthonormal triad $\{\vec{\widehat{e}}_1(s), \vec{\widehat{e}}_2(s),
\vec{\widehat{e}}_3(s)\}$, where $\vec{\widehat{e}}_3$ is tangent to
the curve, while $\vec{\widehat{e}}_1$ and $\vec{\widehat{e}}_2$ lie on
the plane of the ideal, planar Watson-Crick base pairs~\cite{mark94},
with $\vec{\widehat{e}}_1$ directed along the symmetry axis of the two
grooves, pointing in the direction of the major groove.  Orthogonality
then determines $\vec{\widehat{e}}_2 = \vec{\widehat{e}}_3 \times
\vec{\widehat{e}}_1$ (see Fig.~\ref{fig:scheme}).

The three-dimensional shape of the space curve is fully described by the 
3-vector field $\vec{\Omega}$ that rotates the local unit vectors,
\begin{equation}
\frac{\dd\vec{\widehat{e}}_i}{\dd s} = 
\left(\vec{\Omega}+\omega_0 \vec{\widehat{e}}_3 \right) 
\times \vec{\widehat{e}}_i,
\label{eq:rotation}
\end{equation}
where the index $i$ runs over the three spatial directions, and where
$\omega_0$ is the intrinsic twist-density of the double helix.  As is
familiar from mechanics, the rotation vector $\vec{\Omega}(s)+\omega_0
\vec{\widehat{e}}_3$ relates the triad at $s+\dd s$ to that at $s$.
The three components of $\vec{\Omega}(s)$ along the triad axis are
$\Omega_i (s) \equiv \vec{\Omega} \cdot \vec{\widehat{e}}_i(s)$.
$\Omega_1$ and $\Omega_2$ are bending densities (corresponding to the
``tilt'' and ``roll'' deformations, respectively, of the DNA literature),
with the usual curvature of the backbone given by $\kappa \equiv (
\Omega_1^2 + \Omega_2^2)^{1/2}$.  $\Omega_3$ is the twist density, or,
more precisely, the ``excess'' twist over that of the double helix ground
state, $\omega_0$.

Assuming the ground state to be a straight configuration with
constant twist density $\omega_0$, one can interpret $\vec{\Omega}$
as a strain-field associated with a free energy density.  Taking the
symmetries of the double helix into account, the deformation free energy
to second order in $\vec{\Omega}$ is~\cite{mark94}
\begin{equation}
\beta E = \frac{1}{2}\int_0^L \left(
A_1 \Omega_1^2 + A_2 \Omega_2^2 + 
C \Omega_3^2 + 
2G \Omega_2 \Omega_3
\right) \dd s,
\label{eq:energy_gen}
\end{equation}
where $\beta=1/k_BT$ is the inverse temperature, and $A_1$, $A_2$,
$C$ and $G$ are the stiffness parameters.  Equation~\eqref{eq:energy_gen}
is characterized by a twist-bend coupling term connecting a bending
deformation towards the DNA groove ($\Omega_2$) to a twist deformation
($\Omega_3$). $G$ denotes the twist-bend coupling constant, without 
which ($G=0$) one recovers the TWLC.

We investigate the lowest-energy configuration of a circularly-bent DNA 
molecule, a constraint which can be mathematically imposed
by appropriate Lagrange multipliers. This is usually performed by
parametrizing $\Omega_i$ in a lab frame using Euler angles (see e.g.\
Refs.~\cite{bala99, noro08}), and numerically solving the corresponding
Euler-Lagrange equations.  We will instead introduce an approximation,
which will allow us to work in the material frame using the $\Omega$'s
as minimization variables, and perform the minimization analytically.

One might be tempted to fix the curvature $\kappa = (\Omega_1^2 +
\Omega_2^2)^{1/2}$ using a Lagrange multiplier, but this leads to a
helical solution, rather than a closed configuration~\cite{suppl}. This
is a consequence of the bending anisotropy ($A_1 \neq A_2$), together
with the fact that the plane on which the bending takes place is not
restricted.  Instead, we seek to impose bending on a plane, as e.g.\
illustrated in Fig.~\ref{fig:scheme} (left). The bending component of
a local deformation is described by the vector $\vec{\Omega}_b \equiv
\Omega_1 \vec{\widehat{e}}_1 + \Omega_2 \vec{\widehat{e}}_2$.  Enforcing
bending along a fixed plane, as for instance the plane orthogonal to a
vector $\vec{\widehat{x}}$, is equivalent to requiring $\vec{\Omega}_b$
to be parallel to $\vec{\widehat{x}}$. The term $\mu \vec{\Omega}_b
\cdot \vec{\widehat{x}}$ provides a suitable constraint, with $\mu$ as
the Lagrange multiplier. This can be rewritten in the following form
\begin{equation}
\beta \widehat{E} \equiv 
\beta E - \mu \int_0^L \left[ \Omega_1 \sin \phi(s) + 
\Omega_2 \cos \phi(s) \right] \dd s,
\end{equation}
where we have assumed that $\vec{\widehat{x}}$ lies on the plane spanned
by $\vec{\widehat{e}}_1$ and $\vec{\widehat{e}}_2$, and that $\phi$ is
the angle formed between $\vec{\widehat{x}}$ and $\vec{\widehat{e}}_2$
(see Fig.~\ref{fig:scheme}). For a straight DNA lying on the
plane orthogonal to $\vec{\widehat{x}}$ we have $\phi(s) = \omega_0 s$.
If within one helical turn bending is relatively weak (i.e.\ $\kappa
\ll \omega_0$), we can approximate $\phi(s) \approx \omega_0 s$, with
the energy minimization then leading to the simple result
\begin{equation}
\Omega_1 = \frac{{\mu} \sin (\omega_0 s)}{A_1}, \ \ \ \ 
\Omega_2 = \frac{{\mu} \cos (\omega_0 s)}{A_2-G^2/C}, \ \ \ \ 
\Omega_3 = -\frac G C \, \Omega_2,
\label{omega_analytic}
\end{equation}
with ${\mu} \equiv l_b/R$, where $R$ is the average radius
of curvature and $l_b$ the bending persistence length of the
model~\eqref{eq:energy_gen} \cite{nomi17}. The Supplemental
Material~\cite{suppl} discusses the details of the calculations and
alternative approaches~\cite{moha05b}.

The equations~\eqref{omega_analytic} describe a curve with small
off-planar periodic fluctuations appearing in the form of standing
waves in bending and twist. A non-vanishing $G$ is essential for the
emergence of twist waves~\footnote{Ref.~\cite{olso04} (based upon
the theory developed in~\cite{cole03}) analyzes tightly circularly
bent DNA numerically, for a model that includes twist-bend coupling of
precisely the form introduced in~\cite{mark94}. In accord with analytical
predictions of~\cite{mark94}, the numerical results of~\cite{olso04}
for a homogeneous DNA model display twist waves (see Fig. 3, lower
panel). However, \cite{olso04} does not discuss the correlation
between local bending and twisting as a generic feature of all DNA
sequences, nor does it analyze tight DNA bending and twist waves in
experimental data from nucleosome crystal structures (Fig. 3 of this
paper).}. Although our minimization is not exact, as it is performed
under a fixed ``background'' $\phi(s)$, simulations of DNA minicircles of
radii $\approx 5$~nm (see below, \cite{suppl}) are in excellent agreement
with Eq.~\eqref{omega_analytic}. In an alternative approach~\cite{suppl}
one can obtain twist-waves using a systematic perturbation scheme in
powers of $\kappa/\omega_0$, similar to that of Ref.~\cite{mark94}; this
parameter is $\kappa/\omega_0 \approx (1/5)/1.75 \approx 0.11$ for a DNA
minicircle of radius $5$~nm, justifying our approximation~\cite{suppl}.

\begin{figure}[t!]
\centering\includegraphics[width=\linewidth]{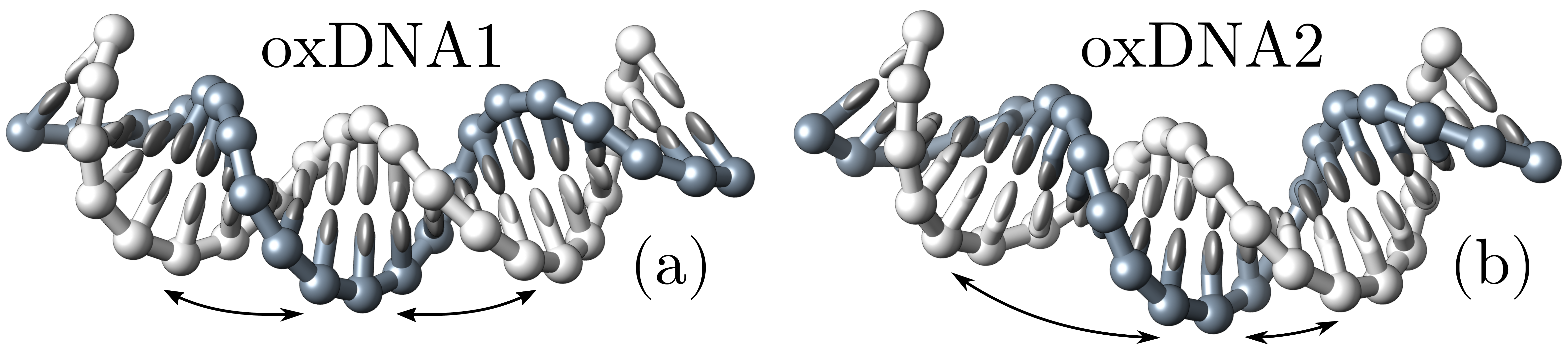}
\caption{(a,b) Snapshots of minicircles fragments from simulations of
oxDNA1 (with symmetric grooves, (a)) and of oxDNA2 (with asymmetric
grooves, (b)).}
\label{fig:oxDNA}
\end{figure}

\begin{figure*}[t!]
\includegraphics[width=\linewidth]{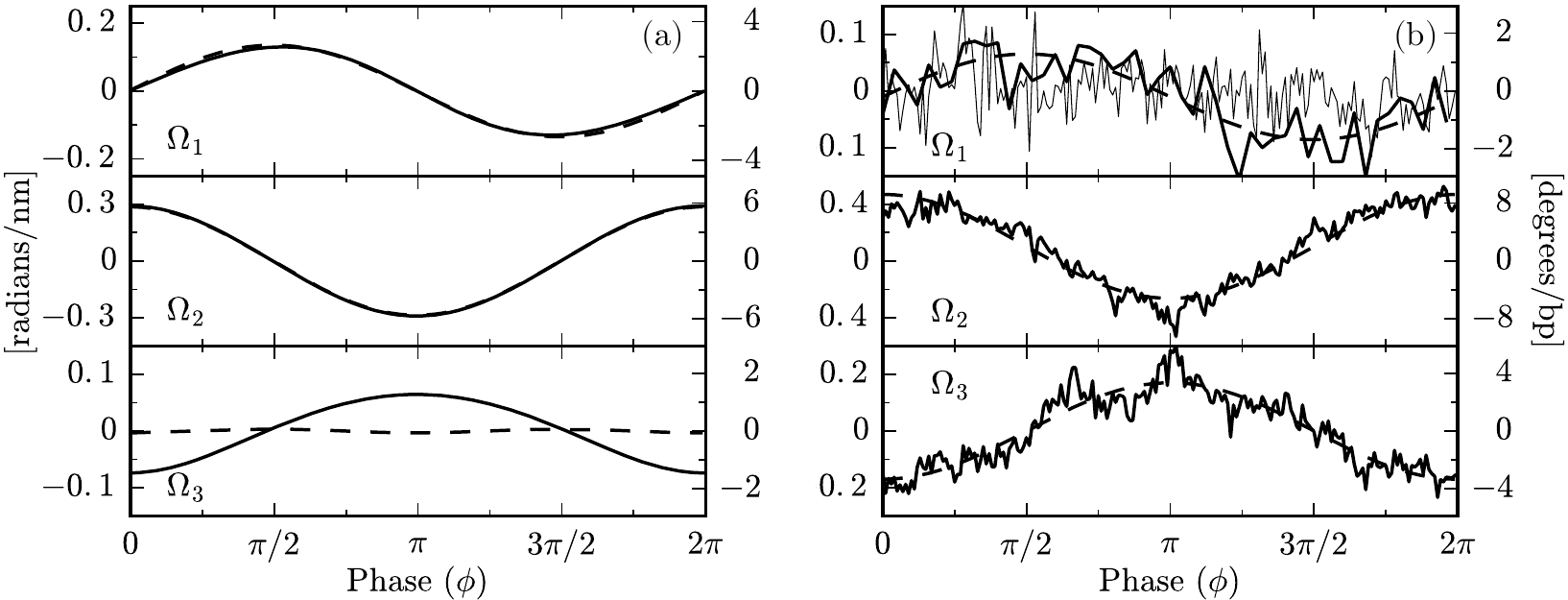}
\caption{
(a) Plot of average values of $\Omega_i$ vs.\ $\phi$ from oxDNA1 (dashed
lines) and oxDNA2 (solid lines) simulations. oxDNA2, but not oxDNA1,
has a pronounced twist wave. Overall the data are in good agreement with
Eqs.~\eqref{omega_analytic}.  A zoom-in of the $\Omega_3$ for oxDNA1 shows
a very weak wave with frequency $2\omega_0$. This is due to anisotropic
bending, as discussed in the Supplemental Material~\cite{suppl}.
The $\Omega_i$, as defined in~\eqref{eq:energy_gen}, have units of inverse
length, which are shown in the left vertical axis. The right axis is in
degrees per base pairs, and is obtained by multiplying the left scale
by $180 a /\pi$, with $a=0.34$~nm the base pair separation.  (b) Plot
of the mean values of $\Omega_i$ vs.\ the phase $\phi$ (analogously to
Fig.~\ref{fig:Omegas}), obtained from averaging over $145$ nucleosome
crystal structures.  Noisy curves for $\Omega_2$ and $\Omega_3$ are
simple averages over all structures; smooth curves show the Fourier
component for $\omega_0$, indicating its dominance in the average, as
well as the antiphase relation of $\Omega_2$ and $\Omega_3$ expected
from the twist-bend coupling.  Data for $\Omega_1$ averaged over all
structures are extremely noisy (light noisy curve), but when selected
structures with large power at $\omega_0$ are analyzed (darker curves)
the $\pi/2$-phase-shifted signal expected from theory is observed (see
text). The output of the software Curves+~\cite{lave09} is in degree
per bp, given in the right vertical axis.}
\label{fig:Omegas}
\end{figure*}

{\sl Coarse-grained DNA simulations --} 
We have performed computer simulations of minicircles with oxDNA, a
coarse-grained DNA model in which the double helix is composed of two
intertwined strings of rigid nucleotides, held together by non-covalent
interactions~\cite{ould10, snod15}. Base-pairing together with all
other interactions are homogeneous, i.e.\ sequence-dependent effects
are neglected.  Various aspects of the mechanics of DNA minicircles,
such as kinking, melting and supercoiling, have been discussed in
the literature using oxDNA, other coarse-grained models or all-atom
simulations~\cite{amza06, fogg06, lank06, noro08, sutt16}.  Here we focus
on the ground-state shape of homogeneous minicircles, and in particular
on circular molecules of $85$~base pairs (bp), or about $29$~nm in length
(see Fig.~\ref{fig:scheme}). With this choice of length the two ends of
the molecule can be joined together without introducing an excess linking
number. In addition, the radius of the circles $R=4.6$~nm is close to
that of nucleosomal DNA ($R=4.2$~nm) which will be analyzed later. Two
versions of oxDNA were used, see Fig.~\ref{fig:oxDNA}(a,b). In the first
version (oxDNA1) the helical grooves have equal width~\cite{ould10},
while in the second version (oxDNA2) the grooves are asymmetric, as
in real DNA~\cite{snod15}. More details on simulations can be found in
Supplemental Material~\cite{suppl}.

Figure~\ref{fig:Omegas}(a) shows a comparison between oxDNA1 and
oxDNA2 simulations (dashed and solid lines, respectively), in which
the $\Omega_i$ are plotted as a function of the base-pair phase angle
$\phi$. The latter was obtained from a Fourier analysis of simulation
data: a discrete Fourier transform provides a dominant frequency
$\omega_0$ and a global phase $\psi$. From these the local phase of
each individual base pair was obtained as $\phi_n = \text{mod} (\psi +
n a \omega_0,2\pi)$, with the index $n=0, 1\ldots 84$ labeling the base
pairs along the circle, and $a=0.34$~nm being the base pair separation.
The smooth curves of Fig.~\ref{fig:Omegas}(a) are obtained by binning the
data in $\phi$ and averaging $\Omega_i$ within each bin. A key result
of Fig.~\ref{fig:Omegas}(a) is the clear difference in the behavior of
$\Omega_3$ between the model with symmetric grooves (oxDNA1, dashed lines)
and that with asymmetric grooves (oxDNA2, solid lines). The emergent twist
waves are associated with the twist-bend coupling interaction [$G \neq 0$
in Eqs.~\eqref{omega_analytic}], which arises from the groove asymmetry
of DNA~\cite{mark94}.  In the unrealistic case of equal major and minor
grooves, one expects $G=0$, as we indeed observe for oxDNA1. In general,
the $\Omega_i$ calculated from oxDNA closely follow the predictions
of Eqs.~\eqref{omega_analytic}. For a quantitative comparison see
Supplemental Material~\cite{suppl}.

{\sl Nucleosomal DNA --} 
We now turn to the analysis of nucleosomal DNA, which is highly bent
around histones, forming a superhelix of radius $4.19$~nm and pitch
$2.59$~nm (for a recent review see e.g.\ Ref.~\cite{esla16}). The length
of the wrapped DNA is $147$~bp, corresponding to $1.67$ superhelical
turns. High-resolution structural crystallographic data for DNA wrapped
around histone proteins in nucleosomes is available (we note the
seminal work of this type in Ref.~\cite{rich03}).  Oscillations in tilt
($\Omega_1$), roll ($\Omega_2$) and twist ($\Omega_3$) were found in
early analyses of crystallographic data, and were attributed
to histone protein-DNA interactions~\cite{rich03}. Since the
publication of the first high-resolution nucleosome data~\cite{rich03},
many crystal structures have been determined with different wrapping
sequences and various DNA or protein modifications (e.g. methylation and
phosphorilation).  Here we focus on the average shape of nucleosomal DNA,
which can be obtained by averaging over different available structures.
Nucleosomal DNA forms a superhelix and not a close circle.  Nonetheless,
Eqs.~\eqref{omega_analytic} are expected to approximate well its shape,
as the superhelical pitch is small compared to the intrinsic double-helix
twist (details in Supplemental Material~\cite{suppl}, see also 
Ref.~\cite{moha05}).

Figure~\ref{fig:Omegas}(b) shows a plot of average $\Omega_i$ vs.\
$\phi$, extracted from the analysis of $145$ crystal structures
from the Protein Data Bank (PDB~\cite{pdb}), using the conformational
analysis software Curves+~\cite{lave09}. The phase $\phi$ is calculated
from the discrete Fourier analysis, similarly to the oxDNA data of
Fig.~\ref{fig:Omegas}(a). From the analysis of crystal structures
we find that in nucleosomal DNA $\Omega_2$ and $\Omega_3$ have a
strong oscillatory behavior for all sequences and are in antiphase as
predicted by Eqs.~\eqref{omega_analytic}. The average of $\Omega_1$
over all crystallographic data results in a structureless, highly-noisy
signal (thin lines, top of Fig.~\ref{fig:Omegas}(b)). However,
a subset of data ($24$ PDB entries out of the $145$ analyzed) show
oscillations in $\Omega_1$, detectable from a dominant peak in the
Fourier spectrum corresponding to a frequency $\approx \omega_0$. The
average of this oscillating subset is a sinusoidal wave, as expected
from Eq.~\eqref{omega_analytic}. The lack of a clear oscillatory signal
may be due to sequence-specific effects and low signal-to-noise ratio,
masking the expected behavior.

There is a reasonable quantitative agreement in the wave amplitudes
between oxDNA simulations and nucleosome data, as seen by comparing
the vertical scales of Fig.~\ref{fig:Omegas}(a) and (b).  According to
Eqs.~\eqref{omega_analytic} the wave amplitudes depend on the value of the
elastic constants, which may be somewhat different between real DNA and
oxDNA.  Nucleosomal DNA has a larger amplitude in $\Omega_2$ and smaller
in $\Omega_1$ than oxDNA. As shown in Supplemental Material~\cite{suppl},
from Eqs.~\eqref{omega_analytic} it follows that $\max\{ \Omega_1\} +
\max\{ \Omega_2\} = 2/R$, a geometric stiffness-independent constant, $R$
being the radius of curvature.  Using this relation we find $R=4.7$~nm
both for oxDNA1 and oxDNA2, which agrees with the expected radius $R =
85 a/2 \pi = 4.6$~nm for a $85$-bp minicircle.  For the nucleosome,
we obtain $R = 4.5$~nm, which, considering the large uncertainty on
$\Omega_1$, is reasonably close to the known nucleosomal-DNA radius
$R=4.2$~nm.  While the sum of the amplitudes $\Omega_1$ and $\Omega_2$ is
constrained by the geometry, this is not the case for $\Omega_3$. Its
amplitude is larger for the nucleosomal data (Fig.~\ref{fig:Omegas}(b))
than for oxDNA2 (Fig.~\ref{fig:Omegas}(a)), suggesting that oxDNA2 has a
twist-bend coupling constant lower than that of real DNA, in agreement
with a previous analysis~\cite{skor17}.  From the ratio between the
amplitudes of $\Omega_3$ and $\Omega_2$ in Fig.~\ref{fig:Omegas}(b)
and Eq.~\eqref{omega_analytic} we estimate $G/C \approx 0.46$. Recent
analysis~\cite{nomi17} of single-DNA magnetic tweezers experiments on
$7.9$~kbp DNA molecules estimated $G=40(10)$~nm and $C=110(5)$~nm, which
would yield $G/C=0.36(09)$. Although these two ratios are consistent,
some caution is required in their comparison. Simulations have shown
that elastic constants for deformations at the base-pair level, relevant
for the nucleosome, are generally smaller than asymptotic stiffnesses
which are obtained for segments of 10-20 base-pairs, relevant for the
tweezers data~\cite{skor17}.

Elastic rod models have been used in
the past to investigate various features of
nucleosomes~\cite{moha05,tols07,vail07,moro09,beck09a,free14,noro14}. In
particular, the structure of nucleosomal DNA has been
addressed~\cite{moha05} using a model including, besides twist-bend
coupling, a stretching modulus and twist-stretch coupling. The elastic
energy was minimized while keeping the twist density fixed to the
experimentally determined values of Ref.~\cite{rich03}, in order to mimic
the interaction of DNA with the histone-proteins. In Ref.~\cite{noro14}
minimization of a sequence dependent model was performed, while fixing
the base pair orientation in $14$ known DNA-histones interaction
sites~\cite{hall09}.  While partially-constraining the conformation of
the nucleosomal DNA along the sequence allows for sharper predictions
about its local and sequence-dependent behavior, it may obscure some
global features. In particular, our work shows that twist oscillations are an 
intrinsic feature of bent DNA, rather than an explicit consequence of 
DNA-protein interactions.

{\sl Conclusion --} Summarizing, we have shown that in a coarse-grained
model of DNA with asymmetric grooves a bending deformation induces
an oscillating excess twist having the form of a standing wave. We
devised an approximated energy-minimization scheme, which provides
analytical predictions for the shape of bending and twist waves. These
are in excellent agreement with the numerical simulations, and show
that the induced twist waves have a spatial frequency $\omega_0$,
the intrinsic DNA twist-density, and an amplitude which is governed
by the radius of curvature and the DNA elastic constants. We also
showed that crystallographic X-ray nucleosomal DNA data match our
prediction of bend-induced twist waves.  In nucleosomes, oscillations
in DNA twist and bending are usually attributed to the DNA-protein
interactions~\cite{rich03}, but our work shows that twist waves are
general features of bent DNA. We expect that the same kind of correlation
will be observed in other protein-DNA complexes, since twist-bend coupling
is a fundamental physical property of the double helix.

\begin{acknowledgments}
ES acknowledges financial support from KU Leuven Grant No.\ IDO/12/08,
and SN from the Research Funds Flanders (FWO Vlaanderen) grant VITO-FWO
11.59.71.7N.  JM acknowledges financial support from the Francqui
Foundation (Belgium), and from the US NIH through Grants R01-GM105847,
U54-CA193419 and U54-DK107980.
\end{acknowledgments}


%
\end{document}